\documentclass[12pt]{iopart}
\usepackage{iopams}
\usepackage{graphicx}
\newcommand{\be}{\begin{equation}}
\newcommand{\ee}{\end{equation}}
\begin{document}

\title{Casimir energy and the superconducting phase transition.}

\author{ Giuseppe Bimonte \footnote{Talk given by this author.}, Enrico Calloni,  Giampiero Esposito and Luigi Rosa$\|$}

\address{ INFN, Sezione di Napoli, and Dipartimento di
Scienze Fisiche, Complesso Universitario di Monte
S. Angelo, Via Cintia, Edificio N', 80126 Napoli, Italy}

\begin{abstract}
We study the influence of Casimir energy on the critical field of
a superconducting film, and we show that by  this means it might
be possible to directly measure, for the first time, the variation
of Casimir energy that accompanies the superconducting transition.
It is   shown that this novel approach may also help clarifying
the long-standing controversy on the contribution of TE zero modes
to the Casimir energy in real materials.
\end{abstract}

\section{Introduction}

The last ten years have witnessed an intense experimental work on
the Casimir effect \cite{decca}. The terrific improvements in
experimental techniques made it possible to  measure the Casimir
force with an unprecedented  precision, at the level of the
percent, with respect  to the   historical experiments performed
only a few decades ago. It seems fair enough to summarize the
present situation by saying that the experiments on the Casimir
force have shown   good quantitative agreement with the theory,
within the limits that are reasonable for experiments dealing with
macroscopic  physics, and hence one may wonder what comes next.
While there remain important issues to be addressed (most notably
that of thermal corrections in real materials)  which require
further experimental refinements especially at large separations,
we think  the time has come to search for entirely new directions
of experimental activity on the physical effects of vacuum
fluctuations, going beyond force measurements. It occurred to us
that
no experiments yet exist,
which probe
directly the physical effects of Casimir {\it energy}. Energy is a
more fundamental quantity than force, and therefore it seems to us
that  this would be a rewarding target.

In view of the important r$\hat{\rm o}$le that the energy of
vacuum fluctuations may have played in the Early Universe, we
considered two possible directions as interesting.
One deals with the gravitational effects of the Casimir energy,
and indeed some time ago \cite{calloni} we studied the feasibility
of an experiment aimed at verifying the validity of the
Equivalence Principle of General Relativity for the zero-point
energy of vacuum fluctuations. While  we are still working on this
problem,  the findings in \cite{calloni} indicate that such an
experiment might be feasible, provided that signal modulation
problems can be solved. The second direction that we undertook
concerns the influence of Casimir energy on phase transitions. We
studied in particular the superconducting phase transition
\cite{bimonte}, and this contribution provides a summary of the
work done so far. The results are very encouraging, and indeed the
INFN has recently sponsored our experiment ALADIN2, to test the
effects that  are described in this paper. This represents a new
approach to the Casimir effect, that might contribute also to
clarify some controversial issues regarding the Casimir effect in
real materials.

The plan of the paper is as follows: in Sec. 2 we present the
general theoretical ideas involved in our  experiment \footnote{The experimental
aspects of the setup  are discussed in a separate paper in this issue.}, while Sec. 3  explains how
to use Lifshitz theory to compute the variation of Casimir energy
across the superconducting transition. In Sec. 4 we  present the
results of our numerical computations, and in Sec. 5 we examine
the issue of the contribution of the TE zero mode. Finally, Sec. 6
contains our conclusions and a discussion of the results.

\section{The Casimir effect in a superconducting cavity}

Consider the double cavity shown in Fig. 1, obtained by placing a
thin superconducting film, with thickness $D$, between the plates
of a rigid plane-parallel Casimir cavity. The two gaps at the
sides of the film, of common width $L$, are filled with some
insulator.
\begin{figure}
\includegraphics{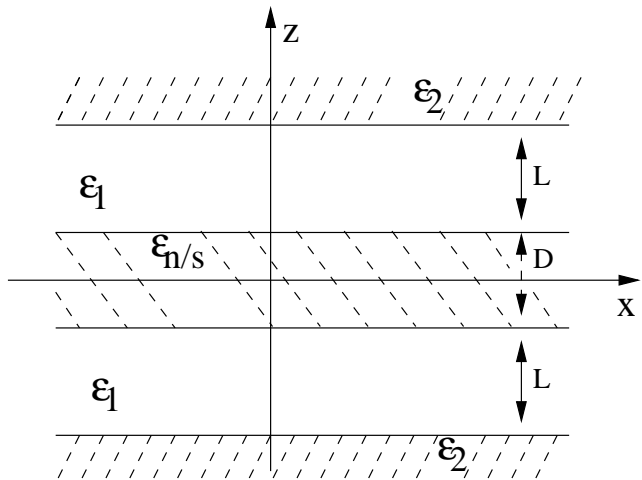}
\caption{\label{fig1} Scheme of the superconducting five-layer
cavity.}
\end{figure}
It is well known that the magnitude of the Casimir effect depends
on the reflective power of the layers forming the cavity. Now,
experiments show \cite{glover} that the reflective properties of a
superconducting film, in the microwave region of the spectrum, are
drastically different from those in the normal  state. Therefore
one can foresee that both the Casimir free energy stored in the
cavity and the Casimir force on the outer plates change when the
film passes from normal ($n$) to superconducting ($s$), and one
wonders if there is a way to measure these effects. A standard
force measurement on the outer plates would be certainly
impractical, because the variation of the Casimir force across the
transition is extremely small \cite{trunov}. The reason for this
is easy to understand, and is due to the fact that the transition
to superconductivity affects the reflective power of the film only
at wavelengths of order $c \hbar/(k T_c) \simeq 2$ mm (for a
typical critical temperature $T_c \simeq 1$ K), which are very far
from the submicron range,  that gives the  dominant contribution
to the Casimir force for typical Casimir cavities. In fact, we
estimated that in typical conditions the relative variation of the
Casimir force is of order $10^{-8}$ or less, which is clearly
unmeasurable within the present level of precision, which is only
of a few percent. Therefore, one has to consider alternative
effects, and we realized that a feasible scheme involves the
measurement of the critical magnetic field required to destroy the
superconductivity of the film. Let us see why this new approach
may very well work.

As is well known, superconductors show perfect diamagnetism, as
they expel magnetic fields from their interior. However, this is
true only for magnetic fields not exceeding a critical value
$H_c$, above which it becomes energetically convenient for the
superconductor to  revert to the normal state and let the magnetic
field in. For standard samples, the value of $H_c$ can be obtained
by equating the magnetic work $W$, done to expel the critical
field, to the so-called condensation energy of the superconductor
${\cal E}_{\rm cond}(T)$, defined as the difference among the
Helmoltz free energies of the film, in the $n$ and in the $s$
state \cite{tink}. For a thick film (with thickness $D$ much
greater than the superconductor penetration depth $\lambda$ and
correlation length $\xi$) and a parallel field, one finds $W=V
H^2/(8 \pi)$, with $V$ the volume of the film, and therefore one
gets for $H_{c\|}$ the equation: \be V\;\frac{H_{c \|}^2}{8
\pi}\;={\cal E}_{\rm cond}(T)\;.\label{hcri} \ee What happens if
the film is placed now inside a Casimir cavity? With respect to
the previous situation, we have to take into account that the
magnetic work $W$ must now be balanced against the condensation
energy of the film {\it plus} the difference $\Delta
F^{(C)}_E(T)=F^{(C)}_n(T)-F^{(C)}_s(T)$ between the Casimir free
energy $F^{(C)}_{n/s}(T)$ in the $n/s$ states of the film,
respectively, and therefore one obtains the following modified
equation for the critical field: \be V\;\frac{ (H_{c \|}^{\rm
cav}) ^2}{8 \pi}\, \,= {\cal E}_{\rm cond}(T)\,+\,\Delta
F^{(C)}_E(T)\;.\label{hcricav}\ee On intuitive ground, we expect
$\Delta F^{(C)}_E(T)$ to be   a positive quantity, because in the
superconducting state the film should be closer to behave as an
ideal mirror, than in the normal state, and therefore
$F^{(C)}_s(T)$ should be more {\it negative} than $F^{(C)}_n(T)$.
In view of Eq. (\ref{hcricav}),  this implies that the critical
field should be shifted by the Casimir term towards larger values.
Upon comparing Eqs. (\ref{hcri}) and (\ref{hcricav}), we see the
shift of critical field should have a relative magnitude
approximately equal to: \be\delta H_{c \|}/H_{c \|} \approx \Delta
F_E^{(C)}(T)/(2{\cal E}_{\rm cond}(T))\;.\label{shift}\ee The key
point to notice here is that the relative shift of critical field
is determined by the ratio of $\Delta F^{(C)}_E(T)$ not to
$F^{(C)}_{n/s}(T)$ (as explained earlier this ratio is going to be
very small indeed) but rather to the condensation energy of the
film. Now, the  latter quantity is very small for a thin film, and
for a film thickness of a few nm \footnote{For such ultrathin
films, Eqs. (\ref{hcri}) and (\ref{hcricav}) above should be
modified to take account of incomplete field expulsion in thin
films. However, the final formula for the relative shift Eq.
(\ref{shift}) remains unaltered. See \cite{bimonte} for details.},
it is easily several orders of magnitude smaller than typical
Casimir energies. Therefore, even if $\Delta F^{(C)}_E(T)$ is a
tiny fraction of the Casimir energy of the cavity, it may be
easily comparable to ${\cal E}_{\rm cond}$, and produce a
measurable shift of critical field. In fact, in the case of a Be
film, we estimated that a relative variation of Casimir energy as
small as one part in $10^8$, could still correspond, close to
$T_c$, to more than $10 \%$ of ${\cal E}_{\rm cond}$, and would
therefore induce a shift of critical field of over $5 \%$!

\section{Computation of $\Delta F^{(C)}_E(T)$}

We have evaluated $\Delta F^{(C)}_E(T)$ by means of Lifshitz
theory for the Casimir effect in dispersive media \cite{lifs}. We
recall that the fundamental physical assumption of that theory is
that one can describe, in the relevant range of frequencies, the
propagation of electromagnetic waves in the material bodies
forming the cavity, by means of a complex permittivity $\epsilon
(\omega)$, depending only on the frequency $\omega$ and not on the
wave-vector ${\bf q}$. Therefore, Lifshitz theory cannot be
applied in situations where space dispersion becomes important. In
our case, the applicability of such a theory to the computation of
$\Delta F^{(C)}_E(T)$ might be questioned, because the
characteristic wavelengths which occur in the determination of
$\Delta F^{(C)}_E(T)$, as pointed out earlier, belong to the
microwave region of the spectrum, where normal metals may show an
anomalous skin effect. This is especially true at cryogenic
temperatures, when the anomalous region further extends, due to
longer electron's  mean free paths. In the superconducting state
of the film, non-local effects may be even more important because,
due to the small skin depth of electromagnetic fields in
superconductors, the anomalous skin effect is observed, in clean
superconductors, even inside the frequency domain characteristic
of the normal skin effect in normal metals (extreme anomalous skin
effect). Fortunately, however, non-local effects are less
important in ultrathin films (with thickness $D$ much smaller than
the penetration depth $\lambda$), than in bulk samples. The reason
is that the electron  mean free paths in ultrathin films cannot be
very large, even in the superconducting state. For example, the
authors of Ref. \cite{adams} quote a  mean free path of $64$ nm in
pure ultrathin superconducting Be films with a thickness $D=4.2$
nm ($T_c=0.6$ K). Therefore, when considering ultrathin films, one
is in the so-called dirty case, where local electrodynamics
remains a valid approximation. This is confirmed by experiments
\cite{glover}, showing that the film conductivity is independent
of film thickness, for small thicknesses.

Let us briefly recall how to compute $\Delta F^{(C)}_E(T)$. As is
well known, there exists a simple derivation of the Lifshitz
formula for the Casimir energy in dispersive media, based on
consideration of the stationary  modes of the cavity \cite{schr}.
This approach is best suited to study  our five-layer system (see
Fig.1), for which the original Lifshitz method would be very
complicated. The electric permittivities of the layers are denoted
as follows: $\epsilon_{n/s}$ represents the permittivity of the
film, in the $n/s$ states respectively, while $\epsilon_1$ is the
permittivity of the insulating layers.  Last, $\epsilon_2$ is the
permittivity of the outermost thick normal metallic plates.
According to the mode method, one can write  the unrenormalized
variation of Casimir energy $\Delta E^{(C)}_0$, at $T=0$, as:
 \be \Delta E^{(C)}_0=\frac{\hbar \, A}{2}  \int
\frac{dk_1 dk_2}{(2 \pi)^2} \sum_{\alpha=TE, TM}  \sum_p
\left(\omega_{{\bf k_\bot},\,p}^{(n,\,\alpha)} \;-\; \omega_{ {\bf
k_\bot},\,p}^{(s,\,\alpha)} \right)\;,\label{unren}\ee where $A
\gg L^2$ is the area,  ${\bf k_\bot}=(k_1,k_2)$ denotes the
two-dimensional wave vector in the $xy$ plane, while $\omega_{{\bf
k_\bot},\,p}^{(n/s,\,TM)}$ ($\omega_{{\bf
k_\bot},\,p}^{(n/s,\,TE)}$) denote the proper frequencies of the
TM (TE) modes,  in the $n/s$ states of the film, respectively.
Upon using the Cauchy theorem  (for details, we address the reader
to Chap. 4 in Ref. \cite{bordag}) \footnote{When comparing the
formulae of this paper with those of \cite{bordag}, please note
that our $L$ and $D$ correspond, respectively, to $d$ and $a$ of
\cite{bordag}, while the TM and TE modes are labelled there by the
suffices (1) and (2), respectively. Note also that in our
configuration the central layer is constituted by the
superconducting film, and not by the vacuum, and hence its
permittivity, denoted by $\epsilon_0$ in \cite{bordag}, is not
equal to 1, but rather to $\epsilon_{n/s}$ depending on the state
of the film.}, we can rewrite   the sums in Eq. (\ref{unren}) as
integrals over {\it complex} frequencies $i \zeta$: \be
\left(\sum_p \omega_{{\bf k_\bot},\,p}^{(n,\,TM)} -\sum_p \omega_{
{\bf k_\bot},\,p}^{(s,\,TM)}\right) =\frac{1}{2
\pi}\int_{-\infty}^{\infty} d \zeta\,   \log
\frac{\Delta^{(1)}_n(i \zeta)}{\Delta^{(1)}_s(i \zeta)}
\;,\label{rensum}\ee where ${\Delta}^{(1)}_{n/s}(i \zeta)$ is the
expression in Eq. (4.7) of \cite{bordag} (evaluated for
$\epsilon_0=\epsilon_{n/s}$). A similar expression can be written
for the $TE$ modes, which involves the quantity
${\Delta}^{(2)}_{n/s}(i \zeta)$ defined in Eq. (4.9) of
\cite{bordag}. It is interesting to note that the integral on the
r.h.s. of Eq. (\ref{rensum}) is {\it finite} because, as observed
earlier, the ratio ${\Delta^{(1)}_n(i \zeta)}/{\Delta^{(1)}_s(i
\zeta)}$  is  appreciably different from one only for frequencies
$\zeta$ in the microwave region (the same is true for the $TE$
contribution). Therefore,  there is no need here for an infinite
renormalization,  contrary to what usually happens when evaluating
Casimir energies. There is however a {\it finite} subtraction to
perform, because we require that the variation of Casimir energy
$\Delta E^{(C)}$ should approach zero for infinite separations
$L$. Upon subtracting from Eq. (\ref{rensum}) (and the analogous
expression for $TE$ modes) its limiting value for $L \rightarrow
\infty$, and after performing the change of variables
$k_\bot^2=(p^2-1)\zeta^2/c^2$ in the integral over $ k_\bot $, one
gets the following expression for the (renormalized) variation of
Casimir energy:  \be \Delta E^{(C)}=\frac{\hbar A}{4 \pi^2
c^2}\int_1^{\infty} p\,dp \int_0^{\infty} d \zeta\,\zeta^2 \,
\sum_{\alpha=TE,TM} \log \frac{Q_n^{\alpha}}{Q_s^{\alpha}} \;,
\label{denren}\ee where \be
 Q_I^{\alpha}( \zeta,p)
=\frac{(1-\Delta_{1I}^{\alpha}\Delta_{12}^{\alpha}e^{-2 \zeta\,K_1
\, L/c})^2 -(\Delta_{1I}^{\alpha}-\Delta_{12}^{\alpha}e^{-2 \zeta
\,K_1\, L/c})^2 e^{-2 \zeta K_I D/c}}{1-(\Delta_{1I}^{\alpha})^2
e^{-2 \zeta K_I \,D/c}}\label{delec} \ee and
\begin{eqnarray}
\Delta_{j\,l}^{TE}=\frac{K_j- K_l}{K_j+K_l}\;\;,\;\;\;\;\;
\Delta_{j\,l}^{TM}=\frac{K_j \,\epsilon_l\,(i \zeta)-K_l
\,\epsilon_j\,(i \zeta)}{K_j\, \epsilon_l\,(i \zeta)+K_l\,
\epsilon_j\,(i \zeta)}\;,\\\;\;K_j=\sqrt{\epsilon_j\,(i
\zeta)-1+p^2}\;,\;\;\;I=n,s\;\;;\;\;j\,,\,l=1,2,n,s.\label{defs}
\end{eqnarray}
The extension of the above formulae to the case of finite
temperature is straightforward. As is well known this amounts to
the replacement in Eq. (\ref{denren}) of the integration $ \int d
\zeta/2 \pi$ by the summation $k T/\hbar \sum_l$ over the
Matsubara frequencies $\zeta_l=2 \pi l/\beta$, where
$\beta=\hbar/(k T)$, which leads to the following expression for
the variation $\Delta F^{(C)}_E(T)$ of  Casimir free energy:
 \be \Delta F^{(C)}_E(T)=A\, \frac{k
\,T}{2}\sum_{l=-\infty}^{\infty}\int \frac{ d {\bf k_\bot}}{(2
\pi)^2} \,\left(\log \frac{Q_n^{TE}}{Q_s^{TE}}+\log
\frac{Q_n^{TM}}{Q_s^{TM}}\right)\;. \label{fint}\ee As we see,
Eqs. (\ref{denren}-\ref{fint}) involve the electric permittivities
$\epsilon\,(i \zeta)$ of the various layers at imaginary
frequencies $i \zeta$. For these functions, we have made the
following choices.

For the outermost metal plates, we use  a Drude model for the
electric permittivity: \be
\epsilon_D(\omega)=1-\frac{\Omega^2}{\omega(\omega+i
\gamma)}\;,\label{drper}\ee where $\Omega$ is the plasma frequency
and $\gamma=1/\tau$, with $\tau$ the relaxation time. We denote by
$\Omega_2$ and $\tau_2$ the values of these quantities for the
outer plates. As is well known, the Drude model provides a very
good approximation in the low-frequency range $\omega \approx 2
k\, T_c/\hbar \simeq 10^{11}\div 10^{12}$ rad/sec which is
involved in the computation of $\Delta F^{(C)}_E(T)$ and $\Delta
E^{(C)}$. The relaxation time is temperature dependent and for an
ideal metal it becomes infinite at $T=0$. However, in real metals,
the relaxation time stops increasing  at sufficiently low
temperatures (typically of order a few K), where it reaches a
saturation value, which is determined by the impurities that are
present in the metal. Since  in a superconducting cavity the
temperatures are very low, we can assume that $\tau_2$ has reached
its saturation value and therefore we can treat it as a constant.
The  continuation of Eq. (\ref{drper}) to the imaginary axis is of
course straightforward and gives \be \epsilon_D(i
\zeta)=1+\frac{\Omega^2}{\zeta\,(\zeta+ \gamma)}
\;.\label{druima}\ee

For the insulating layers, we take a constant permittivity, equal
to the static value:
 \be
\epsilon_1(\omega)=\epsilon_1(0)\;.\ee Again, this is a good
approximation in the range of frequencies that we consider.

As for the film, in the normal state we use  again the Drude
expression Eq. (\ref{drper}),  with appropriate values for the
plasma frequency $\Omega_n$ and the relaxation time $\tau_n$.

The permittivity $\epsilon_s(i \zeta)$ of the film in the
superconducting state cannot be given in closed form and we have
evaluated  it by using the Mattis--Bardeen  formula for the
conductivity $\sigma_s(\omega)$ of a superconductor in the local
limit $q \rightarrow 0$ of BCS theory. Actually, for the
evaluation of $\Delta F^{(C)}_E(T)$ we need only consider the real
part $\sigma'_s(\omega)$ of the complex conductivity, because the
expression of the permittivity $\epsilon_s(i \zeta)$ along the
imaginary axis, which occurs in the Lifshitz formula, can be
obtained from that of $\sigma'_s(\omega)$ by use of the dispersion
relation
 \be \epsilon_s(i \zeta)-1=8
\int_0^{\infty} d\omega \frac{
\sigma'_s(\omega)}{\zeta^2+\omega^2}\;.\label{disp}\ee The reader
can find explicit formulae for $\sigma'_s(\omega)$ in Refs.
\cite{bimonte}. Here, we observe only that $\sigma'_s(\omega)$
can be thought of as the sum of three contributions: a $\delta$
function at the origin, a broad thermal component that diverges
logarithmically at $\omega=0$ and a direct absorption component,
with an onset at $2 \Delta(T)$. At any $T<T_c$, complete
specification of $\sigma'_s(\omega)$ requires fixing three
parameters: besides the free electron density $n$ (or equivalently
the square of the plasma frequency $\Omega_n^2=4 \pi n e^2/m$)
that provides the overall scale of $\sigma'_s$, and the relaxation
time for the normal electrons $\tau_n$, both of which already
occur in the simple Drude formula, $\sigma'_s(\omega)$ only
depends on one extra parameter, i.e. the gap $\Delta$.  We point
out that this expression for $\sigma'_s$ is valid for arbitrary
relaxation times $\tau_n$, i.e. for arbitrary mean free paths, and
in particular it holds in the so-called impure limit
$y=2\Delta/(\hbar \tau_n) \gg 1$, where the effects of
non-locality become negligible.
\begin{figure}
\includegraphics{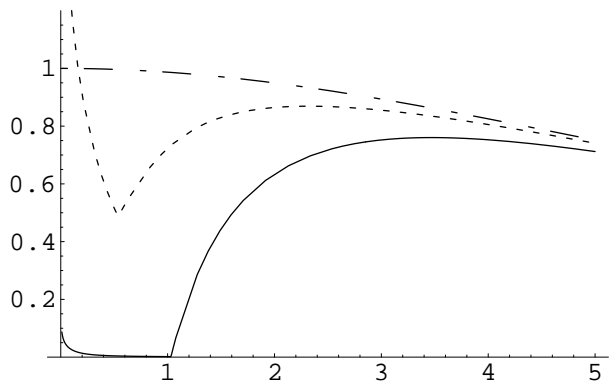}
\caption{\label{plotssigma} Plots of $   m \,\sigma'_s(\omega)/( n
e^2 \tau_n)$, for $T/T_c=0.3$ (solid line), $T/T_c=0.9$ (dashed
line) and $T=T_c$ (point-dashed line). On the abscissa, the
frequency $\omega$ is in   reduced units $x_0=\hbar \omega/(2
\Delta(0))$, and $y_0=2 \Delta(0)/\tau_n \simeq 8.7 $.}
\end{figure}

We  point out that at fixed $\omega$ for $T \rightarrow T_c$, as
well as at fixed $T<T_c$ for $x \equiv \hbar \omega/(2
\Delta)\rightarrow \infty$, $\sigma'_s(\omega)$ approaches the
Drude expression $\sigma_D'(\omega)$ \be
\sigma'_D(\omega)=\frac{1}{4 \pi} \frac{\Omega^2\;
\tau}{1+\omega^2 \tau^2}. \label{drude}\ee The convergence of
$\sigma'_s(\omega)$ to $\sigma_D'(\omega)$ in the frequency domain
is in fact very fast, and already for $x$ of order 10 or so
$\sigma'_s$ becomes undistinguishable from $\sigma_D'$, in
accordance with experimental findings \cite{glover}. In Fig. 2 we
show the plots of $ \sigma'_s(\omega) \,m/ (n e^2 \tau_n)$, for
three values of the reduced temperature $t\equiv T/T_c$, i.e.
$t=0.3, \;0.9$  and $t=1$. The curves are computed for $y_0=2
\Delta(0)/\tau_n \simeq 8.7 $. Frequencies are measured in reduced
units $x_0=\hbar \omega/(2 \Delta(0))$.

\section{Results}

We have evaluated numerically $\Delta F^{(C)}_E(T)$, and the
results of the computation can be summarized as follows:

\begin{itemize}
    \item The contribution of TM modes to $\Delta
    F^{(C)}_E(T)$ is negligible with respect to that of TE
    modes (by three orders of magnitude or so);
    \item $\Delta F^{(C)}_E(T)$ is practically independent (to
    better than four digits) of the value of the dielectric
    constant of the insulating gaps;
    \item $\Delta F^{(C)}_E(T)$ increases with film thickness $D$
    and saturates for $D \simeq c/\Omega_p \simeq 10$ nm;
    \item $\Delta F^{(C)}_E(T)$ increases when the gap separation
    $L$ decreases, and approaches a finite limit, for $L
    \rightarrow 0$;
    \item $\Delta F^{(C)}_E(T)$ increases significantly with the
    plasma frequency of the film $\Omega_n$ and of the outer
    plates $\Omega_2$;
    \item $\Delta F^{(C)}_E(T)$ has a maximum for values around 10
    of the impurity parameter $y$.
\end{itemize}

In Fig. 3 we show the plot of $\Delta F^{(C)}_E$ (in erg) as a
function of the width $L$ (in nm) of the insulating gap, for $D=5$
nm, $T_c=0.5$ K, $y=15$, $t=0.9$, $\Omega_n=\Omega_2=18.9$ eV,
$\tau_2=2.4 \times 10^{-12}$ sec.   We observe that $\Delta
F^{(C)}_E$ is always {\it positive}, which corresponds to the
intuitive expectation that transition to superconductivity of the
film leads to a {\it stronger} Casimir effect, i.e. to {\it lower}
Casimir free energy. The data can be fit very accurately by a
curve of the type \be \Delta F^{(C)}_E(L)  \propto \frac{1}{1+
(L/L_0)^{\alpha}}\;,\ee where $L_0=8.3$ nm and $\alpha=1.15$.
\begin{figure}
\includegraphics{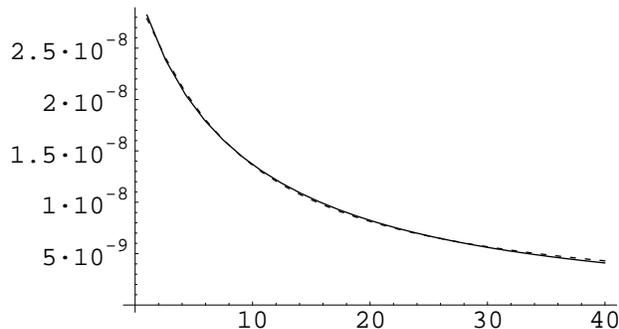}
\caption{\label{figlungo1} Plots of   $\Delta F^{(C)}_E$ (in erg)
as a function of $L$ (in nm)  for $D=5$ nm, $T_c=0.5$ K, $t=0.9$.
See text for the values of the other parameters. Also shown is the
plot (dashed line) of a fit of the type $1/(1+(L/L_0)^{\alpha})$,
with $L_0=8.3$ nm and $\alpha=1.15$.}
\end{figure}

In Fig. 4  we show (solid line) the relative shift of the critical
parallel field of a Be film, as a function of  $t$.   The Figure
has been drawn by using the same parameters as in Fig. 3.
\begin{figure}
\includegraphics{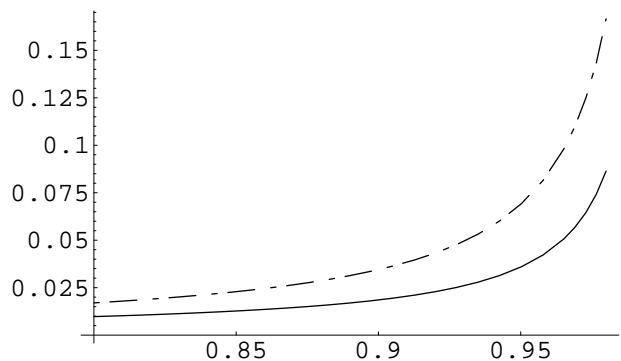}
\caption{\label{figlungo8} Plot (solid line) of  the relative
shift of parallel critical field for a Be cavity, as a function of
$T/T_c$. The point-dashed line has been computed using the plasma
model for the outer plates. Cavity parameters are as in Fig. 3.}
\end{figure}
Note that the shift is {\it positive}, meaning that the critical
field for the film placed in the cavity is {\it larger} than the
critical field for an identical film outside the cavity. The
increase of critical field relative shift as one approaches the
critical temperature  arises because, for $t \rightarrow 1$,
$\Delta F^{(C)}_E$ and ${\cal E}_{\rm cond}$ approach zero at
different rates. Indeed, while ${\cal E}_{\rm cond}$ vanishes as
$(1-t)^2$, $\Delta F^{(C)}_E$ is found to vanish  approximately
like the first power of $1-t$.

\section{Contribution from the $TE$ zero mode.}

In  recent years many efforts have been made to compute the
combined influence of temperature and finite conductivity of the
plates on the Casimir effect,  and no  agreement has been reached
among the experts on the proper way of calculating the
contribution of the $TE$ zero mode (i.e. the $l=0$ term  in the
Matsubara sum) to the Casimir free energy (see Refs. \cite{geyer}
for a   discussion of different points of view on this problem).
This is a delicate issue because, according to Lifshitz theory,
the computation of this mode involves the quantity  \be C:=
 (\zeta^2 \,\epsilon(i \zeta))|_{\zeta=0}
\;.\ee The problem is that, in the Matsubara formalism where
$\zeta$ is discrete,  $C$ is ill-defined if $\epsilon(i \zeta)$
diverges at $\zeta=0$, which is the case for metals, and then the
results depend on how one resolves the ambiguity.   If $\zeta$ is
viewed as a continuous variable,  one may {\it define} $C$ as the
limit of $(\zeta^2 \,\epsilon(i \zeta))$ for vanishing $\zeta$. In
this case, if one uses for the metal conductivity the Drude model
Eq. (\ref{druima}) (with a finite value of $\tau$), one obtains
$C=0$, and this implies that the $TE$ zero mode gives zero
contribution to the Casimir free energy, irrespective of how large
the relaxation time $\tau$ is. The odd thing is that the result is
different if, instead of the Drude model, one uses the simpler
plasma model \be \epsilon(i
\zeta)=1+\frac{\Omega_p^2}{\zeta^2}\;,\ee for then one finds
$C=\Omega_p^2$, and therefore the zero mode gives a non vanishing
contribution, reproducing the ideal metal case in the limit of
infinite plasma frequency.

It is clear that in such a situation it would be very interesting
to have the possibility of an experimental verification of these
effects, and we show below that a superconducting cavity is very
well suited for this purpose.
\begin{figure}
\includegraphics{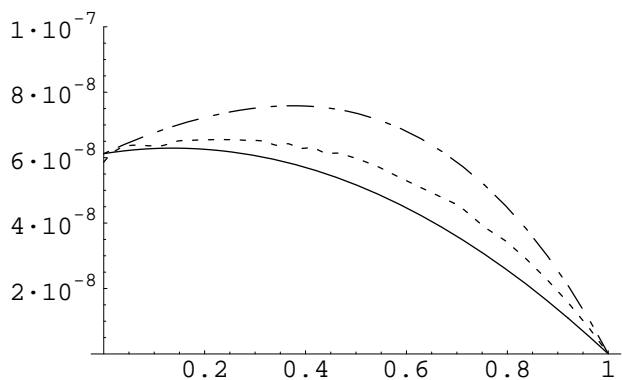}
\caption{\label{figlungo5} Plot (solid line) of $\Delta F^{(C)}_E$
(in erg) as a function of $t$  for $L=10$ nm, $D=5$ nm, $T_c=0.5$
K, $\tau_n=5 \times 10^{-13}$ sec. The point-dashed line was
computed by using the plasma model for lateral plates. Also shown
(dashed line) is the plot of the low-temperature limit of the
Matsubara sum, Eq. (\ref{denren}). See text for further details.}
\end{figure}

The computations presented in the previous Section were performed
by using the Drude model, both for the lateral plates and the film
(in the normal state), and therefore the computed value of $\Delta
F^{(C)}_E$ receives no contribution from the TE zero mode. We have
repeated the computation, by using this time the plasma model  for
the lateral plates and we denote by $\widetilde{\Delta F}^{(C)}_E$
the corresponding value of the variation of Casimir energy. Note
that in this new computation, we keep the Drude model for the film
in the normal state, because  this is the limit of the BCS
expression of the permittivity for $T \rightarrow T_c$. In Fig. 5
we plot (dashed line) $\widetilde{\Delta F}^{(C)}_E$ as a function
of $t$. We point out that the inclusion of the TE   zero mode has
the largest effect  close to $T_c$, where it leads to an
approximate doubling in the value of $\Delta F^{(C)}_E$. The
reason of this is easy to understand: the zero mode becomes more
and more important in the critical region, because a decreasing
number of Matsubara modes contribute to $\Delta F^{(C)}_E$, as one
approaches $T_c$, and therefore the inclusion or omission of a
single mode makes a big difference. That less and less modes
contribute as we move towards $T_c$ is clear, because the
quantities $(Q_{n}^{TE}/Q_{s}^{TE})(i \zeta)$   in Eq.
(\ref{fint}) are substantially different from one only for complex
frequencies $\zeta$ of order a few times $k \,T_c/\hbar$. Since
the $l$-th Matsubara mode has a frequency equal to $2 \pi \,l\,k\,
T/\hbar$, it is clear that the number of terms effectively
contributing to $\Delta F^{(C)}_E$ should be roughly proportional
to $T_c/T$, and hence it is large for $T \ll T_c$, but becomes
small for $T$ comparable to $T_c$. In Fig. 4 we show (point-dashed
line) the shift of critical field resulting from the plasma model,
and we see that the amount of shift is almost doubled with respect
to that  resulting  from use of the Drude model. It seems,
therefore, that if the shift can be detected, it should be rather
easy to distinguish among the two possibilities.

\section{Conclusions and discussion}

We have proposed a novel experimental approach to explore the
physical effects of vacuum fluctuations of the electromagnetic
field, based on the use of superconducting Casimir cavities. In
our scheme, the object of interest is the Casimir energy itself,
rather than the Casimir force, as in all experiments performed so
far. We have shown that the superconducting transition of a thin
film placed between the plates of a plane-parallel cavity,
determines a small variation of Casimir energy. While the
associated variation of the Casimir force on the outer plates is
unmeasurably small, we have found that there is a measurable
effect on the critical magnetic field required to destroy the film
superconductivity.  Because of the Casimir effect, the critical
field is larger than that of a similar film, not placed inside the
cavity. The amount of the shift depends on the temperature, on the
geometric features of the cavity, and on the materials chosen for
the film and for the outer plates, and can be of order a few
percent.

The results presented in this paper represent the initial steps of
a more general experimental research programme, on the influence
of vacuum fluctuations on phase transitions. This is a new
direction in the field of the Casimir effect, that may contribute
to a better understanding of the general issue of the r$\hat{\rm
o}$le of vacuum energy in phase transitions, which is of great
interest in diverse areas of Physics, but most notably in
Cosmology.

We would also like to point out another couple of interesting
features of our approach. One is that we use {\it rigid} cavities,
which may represent an advantage   over conventional Casimir
experiments. As is well known, the experimental difficulty of
controlling the parallelism among macroscopic plane plates with
submicron separations led the experimenters to consider simpler
geometries that do not suffer from this problem, like the
sphere-plane one, which has been adopted in all precision
experiments on the Casimir force (with the only exception of the
experiment by Bressi et al. \cite{decca}, where the plane-parallel
configuration is used, at the price, however, of a reduced
precision compared with the sphere-plate case). This limitation
has made it impossible so far to explore experimentally one of the
distinguishing features of the Casimir effect, i.e. its dependence
on the geometry of the cavity. The use of rigid cavities might
make it possible to study geometries that are difficult to realize
by using non rigid cavities.

Another interesting feature of our scheme relates to the current
interest in the study of the Casimir effect in real materials, in
particular for what concerns the influence of temperature and of
the finite conductivity of the materials. In standard Casimir
force measurements, it is generally quite difficult to measure
these effects, because they typically represent small corrections
to the ideal case, and therefore they are difficult to extract from the
signal. In our  setting, however, the effect  is null if the
film  is treated as an ideal metal, and therefore the signal
arises entirely from the fact that the film is treated as a real
material, with different finite conductivities in the normal and
in the superconducting state. Therefore, our approach seems best
suited to test our understanding of the Casimir effect in real
materials. As an example, we discussed the contribution from the
TE zero mode. This is a controversial issue in the current
literature on thermal corrections to the Casimir effect. It is
well known that, for submicron plate separations, thermal
corrections  to the Casimir force are negligible at cryogenic temperature, and become
relevant  only at room temperature. However, things are different in
our  case, because close to $T_c$ the shift of critical
field is completely determined by the few Matsubara modes with
frequencies below or of order $k T_c/\hbar$, which is where the
reflective properties of a film change when it becomes
superconducting. As a consequence, different treatments of the TE
zero mode lead to strongly different predictions for the shift of
critical field, at the level of doubling the shift,
and this opens the way to a possible experimental clarification of
this delicate problem.

The verification of the effects described in this paper is the
goal of the ALADIN2 experiment, financed by INFN, which is
currently under way at the Dipartimento di Scienze Fisiche
dell'Universit\'a di Napoli Federico II.  Further details on this
experiment can be found in a separate contribution, contained in
this issue.

\section*{Acknowledgments}

We would like to thank A. Cassinese, F. Tafuri, A. Tagliacozzo and
R. Vaglio for valuable discussions, G.L. Klimchitskaya and V.M.
Mostepanenko for enlightening discussions on several aspects of
the problem. G.B. and G.E. acknowledge partial financial support
by PRIN {\it SINTESI}.


\section*{References}

\begin{thebibliography}{99}

\bibitem{decca} Lamoreaux S K 1997 {\it Phys. Rev. Lett.} {\bf 78} 5;  Mohideen U, and Roy A 1998 ibid. {\bf 81}
4549;  Bressi G, Carugno G, Onofrio R, and Ruoso G 2002 ibid. {\bf
88} 041804; Decca R S, L\'opez D, Fischbach E, and  Krause D E
2003 ibid. {\bf 91} 050402;  Chan H B, Aksyuk V A, Kleiman R N,
Bishop D J, and Capasso F 2001 {\it Science} {\bf 291} 1941

\bibitem{calloni} Bimonte G, Calloni E, Di Fiore L, Esposito G, Milano L, and
Rosa L 2006 {\it On the behaviour of a rigid Casimir cavity in a
gravitational field}, in {\it Recent Developments in Gravitational
Physics} Vol. 176 of IOP Conference Series, Eds. I. Ciufolini et
al. (Taylor $\&$ Francis),  hep-th/0302082; Calloni E, Di Fiore L,
Esposito G, Milano L, and Rosa L 2002 {\it Phys. Lett.} A {\bf
297} 328



\bibitem{bimonte} Bimonte G, Calloni E, Esposito G, Milano L, and
Rosa L 2005 {\it Phys. Rev. Lett.} {\bf 94} 180402; Bimonte G,
Calloni E, Esposito G, and  Rosa L 2005 {\it Nucl. Phys.} B {\bf
726} 441



\bibitem{glover}  Glover III R E, and Tinkham M 1957 {\it Phys. Rev.} {\bf
108} 243

\bibitem{trunov} Mostepanenko V M, and Trunov N N 1997 {\it The Casimir Effect and Its
Applications} (Clarendon Press, Oxford)

\bibitem{tink} Tinkham M 1996 {\it Introduction to Superconductivity}
(McGraw--Hill, Singapore)

\bibitem{lifs}  Lifshitz E M 1956 {\it Sov. Phys. JETP} {\bf 2} 73;
 Lifshitz E M, and Pitaevskii L P 1980 {\it Landau and Lifshitz Course of
Theoretical Physics: Statistical Physics Part II}
(Butterworth--Heinemann)

\bibitem{adams} Adams P W, Herron P, and Meletis E I 1998 {\it Phys.
Rev.} B {\bf  58} R2952

\bibitem{schr} Schram K 1973 {\it Phys. Lett.} A {\bf 43} 282

\bibitem{bordag} Bordag M, Mohideen U, and  Mostepanenko V M 2001
{\it Phys. Rep.} {\bf 353} 1

\bibitem{geyer} Geyer B, Klimchitskaya G L, and
Mostepanenko V M 2003 {\it Phys. Rev.} A {\bf 67} 062102; H$\o$ye
J S, Brevik I,   Aarseth J B, and   Milton K A 2003 {\it Phys.
Rev.} E {\bf 67} 056116; Decca R S, Lopez D, Fischbach E,
Klimchitskaya G L, Krause D E, and Mostepanenko V M (2005) {\it
Ann. Phys.} (N.Y.) {\bf 318} 307,  and references therein

\end{thebibliography}
\end{document}